 \documentclass[]{article}

\usepackage{graphicx}

\begin{document}

\title{Attenuation of decameter sky noise during x-ray solar flares in 2013-2017 based on the 
observations at midlatitude radars}

\author{O.I.Berngardt
\and
J.M.Ruohoniemi
\and
N.Nishitani
\and
S.G.Shepherd
\and
W.Bristow
\and
E.S.Miller}




\maketitle

\begin{abstract}
Based on a joint analysis of the data from 10 midlatitude decameter
radars the effects are investigated during 80 x-ray flares in 
the period 2013-2017. For the investigation nine mid-latitude SuperDARN radars
of the northern hemisphere (Adak Island West and East radars, Blackstone radar, 
Christmas Valley East and West radars, Fort Hays East and West radars, 
Hokaido East radar and Wallops radar) and
Ekaterinburg coherent decameter radar of ISTP SB RAS are used. All the radars work in the 
same 8-20MHz frequency band and have similar hardware and software. During the analysis 
the temporal dynamics of noise  from each of the radar direction 
and for each flare is investigated separately. As a result, on the 
basis of about 13000 daily measurements we found a strong anticorrelation 
between noise power and x-ray flare intensity, that allows to use short-wave
sky noise to diagnose the ionospheric effects of x-ray solar flares.
It is shown that in 88.3\% of cases an attenuation of daytime decameter
radio noise is observed during solar flare, correlating with the temporal
dynamics of the solar flare. The intensity of decameter noise anticorrelates
well (the Pearson correlation coefficient better than -0.5) with
the shape of the X-ray flare in the daytime (for solar elevation angle $>0$) in 33\% of
cases, the average Pearson correlation over the daytime is about -0.34.
Median regression coefficient between GOES 0.1-0.8nm x-ray intensity and 
daytime sky-noise attenuation is about $-4.4\cdot10^{4}\ [dB\cdot m^{2}/Wt]$. 
Thus, it has been shown that measurements of the decameter sky noise level
at midlatitude decameter radars can be used to study the ionospheric absorption
of high-frequency waves in the lower ionosphere during x-ray solar flares. This
can be explained by the assumption that the most part of decameter
sky noise detected by the radars can be interpreted as produced by 
ground sources at distances of the first propagation hop (\textasciitilde{} 3000 km).
\end{abstract}

\section{Introduction and motivation}

The widely known impact of solar flares to the Earth's ionosphere is
the increase of the electron density due to additional ionization.
This increase of the electron density in the lower ionosphere (in
particular, at the D-layer heights) often leads to the increase of
absorption of HF and VHF radio waves \cite{Rose_1962,Hunsucker_2002}.
Effect well known as Radio Blackout and widely studied at high latitudes
by riometers \cite{Little_1959,Kikuchi_1989,Kainuma_2001,Hunsucker_2002,Birch_2013}.
The absorption can be described by different models \cite{DRAP2,Rogers_2016},
and most of them includes the following mechanisms: X-ray induced absorption (uses
0.1-0.8nm x-ray GOES data as proxi) and solar energetic particles induced
absorption (uses integral proton fluxes above two energy levels -
5.2MeV and 2.2MeV) \cite{Sauer_2008,DRAP}. At mid-latitudes this
effect is weaker \cite{Hunsucker_2002},
and riometers are used relatively rarely. However, variations of the HF
radio waves absorption at long radio paths including midlatitude ones 
during disturbed conditions are present, intensively investigated \cite{Milan_1996,Breed_2002,Gauld_2002,Eccles_2005,Watanabe_2014,Blagoveshchenskii_2015,Rogov_2015,Blagoveshchenskii_2016} 
and estimated by the different prediction services (for example by NOAA at http://www.swpc.noaa.gov/products/3-day-forecast).

So making monitoring systems for high-frequency absorption effects
at midlatitudes looks an important task, that stimulates extend of 
absorbtion observations to mid-latitudes \cite{Rogers_2016}.

Decameter radars with a narrow antenna pattern are widely used for
the ionospheric studies, at mid- and high latitudes \cite{Greenwald_1995,Chisham_2007,Baker_2007,Kataoka_2007,Vertogradov_2010,Ribeiro_2012,Berngardt_2015}.
The widest network of such radars is the international network of
coherent decameter radars SuperDARN (Super Dual Auroral Radar Network).
Most of these radars are located near the polar caps \cite{Greenwald_1995,Chisham_2007}.
The recent expansion of the network to midlatitudes\cite{Baker_2007,Kataoka_2007,Ribeiro_2012}
makes it possible to use their equipment to diagnose the mid-latitude
effects of solar flares. 

Investigations of the effects of solar flares based on the data from
decameter radars was carried out repeatedly\cite{Watanabe_2014,Squibb_2015}. The main and well known
effect is the strong absorption of the propagating signal during the
flare. During powerful (for example X and sometimes during M) flares
the absorption is so strong that it does not allow us to study in detail 
the dynamics of ionospheric ionization during a solar flare. During such flares
we can only detect the presence or absence of a signal. 
By studies of Doppler velocity of the radar echoes and their range dependence during flares,
estimates of the electron density change rate in the lower ionosphere can be made\cite{Watanabe_2014}.

The use of external sources of radiowaves 
can be used also for diagnosis\cite{Contreira_2005,Squibb_2015}, but it seems not providing a good (world-wide)
spatial coverage, and requires the use of permanent sources of radio
emission for the organization of  world-wide monitoring network.
Sometimes it is not convenient.

Studies of noise level during solar flares have been conducted for
a long time. The main effect during solar flares is increase of the
noise level at almost all radiowave frequencies, that are produced
by Sun \cite{Scherer_2005,Cerruti_2006,Carrano_2009}. This is caused
by the fact that radio frequencies below the gyrofrequency $f_{H}$(of
the order of 1MHz) can propagate
from the space to the Earth's surface by extraordinary mode. Very high frequency waves
with frequencies above 30MHz are also slightly distorted by
the ionosphere and can reach the Earth's surface also \cite{Budden_1985}.
So the main effect of solar flares to the radio noise level is increase
of it. But intermediate frequency range - high frequencies (ranges
between 1-30MHz) has an opposite
properties - the radiowaves propagating from the space can be reflected
by the top part of the Earth's ionosphere and will not reach the Earth's surface.
So the effects of solar flares in the HF bound can differ from the effects
in higher and lower frequency bound. 

Decameter noise usually is divided into 3 parts \cite{Maslin_2003,Bianchi_2007}:
atmospheric (generated mostly by lightings), anthropogenic (generated
by human activity) and cosmic (generated outside the Earth) ones. Above
10MHz the most powerful noise mostly anthropogenic at daytime and mixture
of atmospheric and anthropogneic during nighttime \cite{Maslin_2003,Bianchi_2007}.

The EKB ISTP SB RAS radar is similar by hardware and software with SuperDARN
radars. The EKB radar was launched at the end of 2012 in the Sverdlovsk
Region of the Russian Federation \cite{Berngardt_2015}. Recently,
the data from this radar have been used for the primary study of the
effects of x-ray solar flares in passive mode \cite{Berngardt_2017}.
The results obtained allowed us to assume that the HF sky noise level measured by the radar 
anticorrelates with x-ray flare intensity.

\begin{figure}
\includegraphics[scale=0.35]{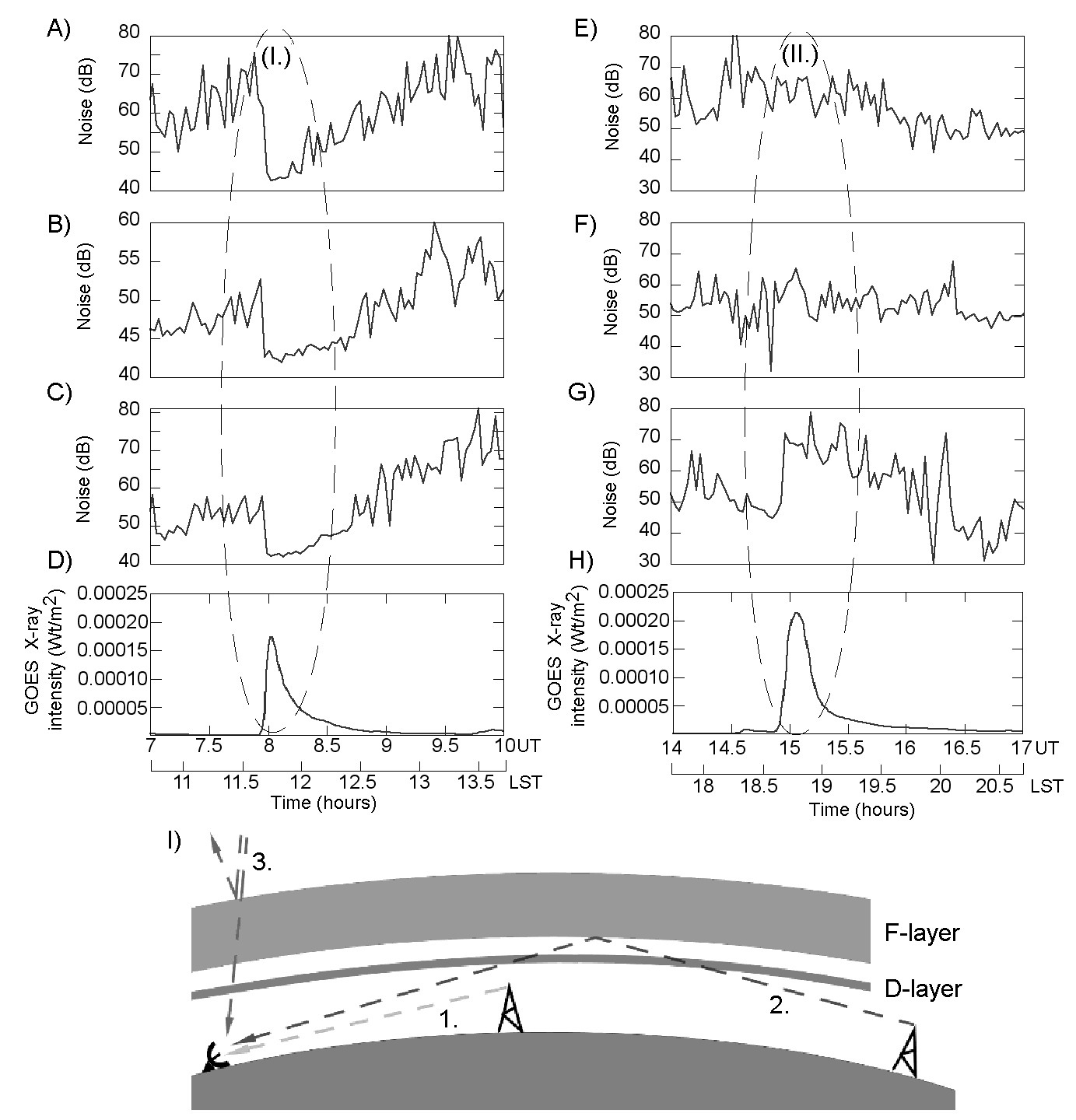}

\protect\caption{A-G) Dynamics of HF sky noise level based on EKB ISTP SB RAS radar data
at different azimuths (beams) during two 25/10/2013 solar flares as a function of universal time and radar local solar time. 
A,E) - beam 15; B,F) - beam 7; C,G) - beam 0. 
Region
I. is the presence of x-ray flare effect; region II. is absence of the
x-ray flare effect. D,H) X-ray intensity dynamics during the flares according to
GOES 0.1-0.8 nm data; I) Illustration of the geometry of the experiment
and an explanation of the trajectories of noise formation and propagation
(1. - ground-wave noise; 2.-sky-wave noise; 3. cosmic noise). }
\label{fig:ekb_noise}\label{fig:ill_geometry}
\end{figure}

In Fig.\ref{fig:ekb_noise} shown a fairly typical behavior of the
daytime decameter sky noise level during a solar flare according to
the EKB radar data at several directions of the antenna pattern (beams).
One can see the decrease of the noise level (Fig.\ref{fig:ekb_noise}A-E) during x-ray flare.
Also one can see that temporal dynamics of the noise anticorrelates with x-ray flare intensity 
measured by GOES mission (Fig.\ref{fig:ekb_noise}F).
As one can see from the figure, not every solar flare
produces this effect (see the regions I. and II. at Fig.\ref{fig:ekb_noise}A-C).

Regular observations of the attenuation effect at EKB radar \cite{Berngardt_2017}
allows us to assume that most part of of the sky noise is produced by terrestrial 
sources, and the most part of the noise comes from sources located at distances 
above the dead zone by sky wave mechanism (because the ground wave at HF range
signifficantly attenuates with the distance). The propagation chractersitics at 
HF allows us also to suggest that the sources producing maximal intensity of the noise 
at the radar are located at the border of dead zone (at the range of groundscatter - the region 
that produces most intensive signal due to focusing of radiowaves in the ionosphere).
So the noise is transmitted by the source, propagated by sky-wave, reflected from the 
ionosphere and absorbed at D-layer altitudes \cite{Hunsucker_2002}. The scheme for the qualitative explanation of the effect
is shown in Fig.\ref{fig:ill_geometry}.

In the paper the above hypothesis of dependence between HF sky noise level
and x-ray solar flare intnesity is verified based on a large statistical 
dataset produced by 10 mid-latitude decameter radars in the northern
hemisphere during 80 x-ray solar flares.

\section{Experimental setup and analysis techniques}

\begin{figure}
\includegraphics[scale=0.6]{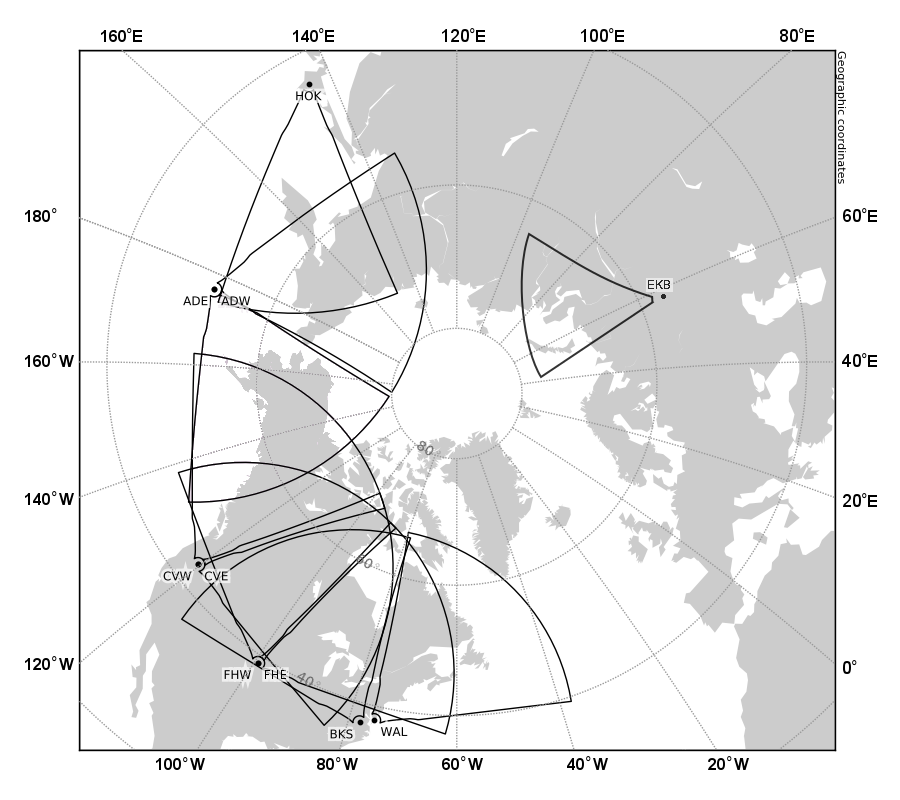}

\protect\caption{Radars used in the paper and their field-of-views}
\label{fig:radars}
\end{figure}

To verify the hypothesis, we used the data of 10 mid-latitude coherent
decameter radars of the northern hemisphere: nine radars of the
SuperDARN network (Super Dual Auroral Radar Network \cite{Chisham_2007})
and the EKB ISTP SB RAS radar \cite{Berngardt_2015}. The coordinates
of the radars and their fields of view are shown in Fig.\ref{fig:radars}.
The basis of the radars operation is transmitting a seria of short
pulses and receiving the signals backscattered by various inhomogeneities
(both ionospheric and terrestrial ones). Each of the radar can operate
at 8-20MHz frequency bound. Each radar can change the direction of main beam 
within its field of view (by changing the beam number - the azimuthal direction 
of the radar antenna pattern).
Each beam width is about 3-6 degrees (depending on the radar frequency),
the radar field of view is about 52 degrees. In addition to the basic
mode of operation, each radar measures the intensity of background
sky noise coming from the given direction. Usually, the radars operate
near a fixed frequency, choosing the exact operation frequency as
frequency with the lowest sky noise level. The radars record the noise
level at the least noisy frequency, allowing subsequent analysis of
the noise level. The regular
temporal resolution of the radars is 1-2 minutes. As it will be shown later this 
temporal resolution is enough for analysis of temporal dynamics of HF sky 
noise during x-ray solar flares, used in this paper.

\begin{table}
\begin{tabular}{|c|c|c|c|}
\hline 
Code & Radar & Coordinates & Owner \tabularnewline
\hline 
ADW & Adak Island West, SuperDARN  & 51.9N,176.6W & Alaska Fairbanks, USA \tabularnewline
ADE & Adak Island East, SuperDARN & 51.9N,176.6W & Alaska Fairbanks, USA \tabularnewline
BKS & Blackstone, SuperDARN & 37.1N,77.9W & Virginia Tech, USA \tabularnewline
CVE & Christmas Valley East, SuperDARN & 43.3N,120.4W & Dartmouth College, USA \tabularnewline
CVW & Christmas Valley West, SuperDARN & 43.3N,120.4W & Dartmouth College, USA \tabularnewline
EKB & Ekaterinburg, ISTP SB RAS & 56.5N,58.5E & ISTP SB RAS,Russia \tabularnewline
FHE & Fort Hays East, SuperDARN & 38.9N,99.4W & Virginia Tech, USA \tabularnewline
FHW & Fort Hays West, SuperDARN & 38.9N,99.4W & Virginia Tech, USA \tabularnewline
HOK & Hokaido East, SuperDARN & 43.5N,143.6E & Nagoya University, Japan \tabularnewline
WAL & Wallops, SuperDARN & 37.9N,75.5W & JHU APL, USA \tabularnewline
\hline 
\end{tabular}

\protect\caption{List of radars, participated in the research}
\label{tab:radars}
\end{table}

For analysis the following technique is used. Based on the GOES 0.1-0.8nm
data, all the x-ray solar flares with intenisty 
exceeded $10^{-5}\, Wt/m^{2}$ during 2013-2017 years were selected. 
If several flashes were observed in the same day we investigate only
the effects of the flare with a maximal amplitude. Using this approach
80 flares were found. The list of investigated days is shown in Table \ref{tab:flares}.

\begin{table}
\begin{tabular}{|c|c|c|c|c|c|}
\hline 
2013-04-11 & 2013-05-03 & 2013-05-10 & 2013-05-13 & 2013-05-14 & 2013-05-15 \tabularnewline
\hline 
2013-06-07 & 2013-10-24 & 2013-10-25 & 2013-10-28 & 2013-10-29 & 2013-11-01\tabularnewline
\hline 
2013-11-03 & 2013-11-05 & 2013-11-06 & 2013-11-08 & 2013-11-10  & 2013-11-19\tabularnewline
\hline 
2013-12-31 & 2014-01-01 & 2014-01-07  & 2014-01-27 & 2014-01-28  & 2014-01-30 \tabularnewline
\hline 
2014-02-02 & 2014-02-04 & 2014-02-25 & 2014-03-12 & 2014-03-29 & 2014-04-02 \tabularnewline
\hline 
2014-04-18 & 2014-04-25 & 2014-05-08 & 2014-06-10 & 2014-06-11 & 2014-07-08\tabularnewline
\hline 
2014-08-24  & 2014-09-10 & 2014-09-28 & 2014-10-02 & 2014-10-16 & 2014-10-19\tabularnewline
\hline 
2014-10-20  & 2014-10-22 & 2014-10-24 & 2014-10-25 & 2014-10-26 & 2014-10-27 \tabularnewline
\hline 
2014-10-28 & 2014-10-30 & 2014-11-03 & 2014-11-05 & 2014-11-06 & 2014-11-07 \tabularnewline
\hline 
2014-11-16 & 2014-12-04 & 2015-01-13  & 2015-03-02 & 2015-03-03 & 2015-03-07 \tabularnewline
\hline 
2015-03-09 & 2015-03-10& 2015-03-11 & 2015-03-12 & 2015-04-21 & 2015-05-05\tabularnewline
\hline 
2015-06-21 & 2015-06-22 & 2015-06-25  & 2015-08-22 & 2015-08-24 & 2015-09-28 \tabularnewline
\hline 
2015-10-01 & 2015-10-02 & 2015-12-23 & 2016-04-18 & 2016-07-23 & 2017-04-01 \tabularnewline
\hline 
2017-04-02 &  2017-04-03 & & & & \tabularnewline
\hline 
\end{tabular}

\protect\caption{List of investigated x-ray solar flares}
\label{tab:flares}
\end{table}

\begin{figure}
\includegraphics[scale=0.6]{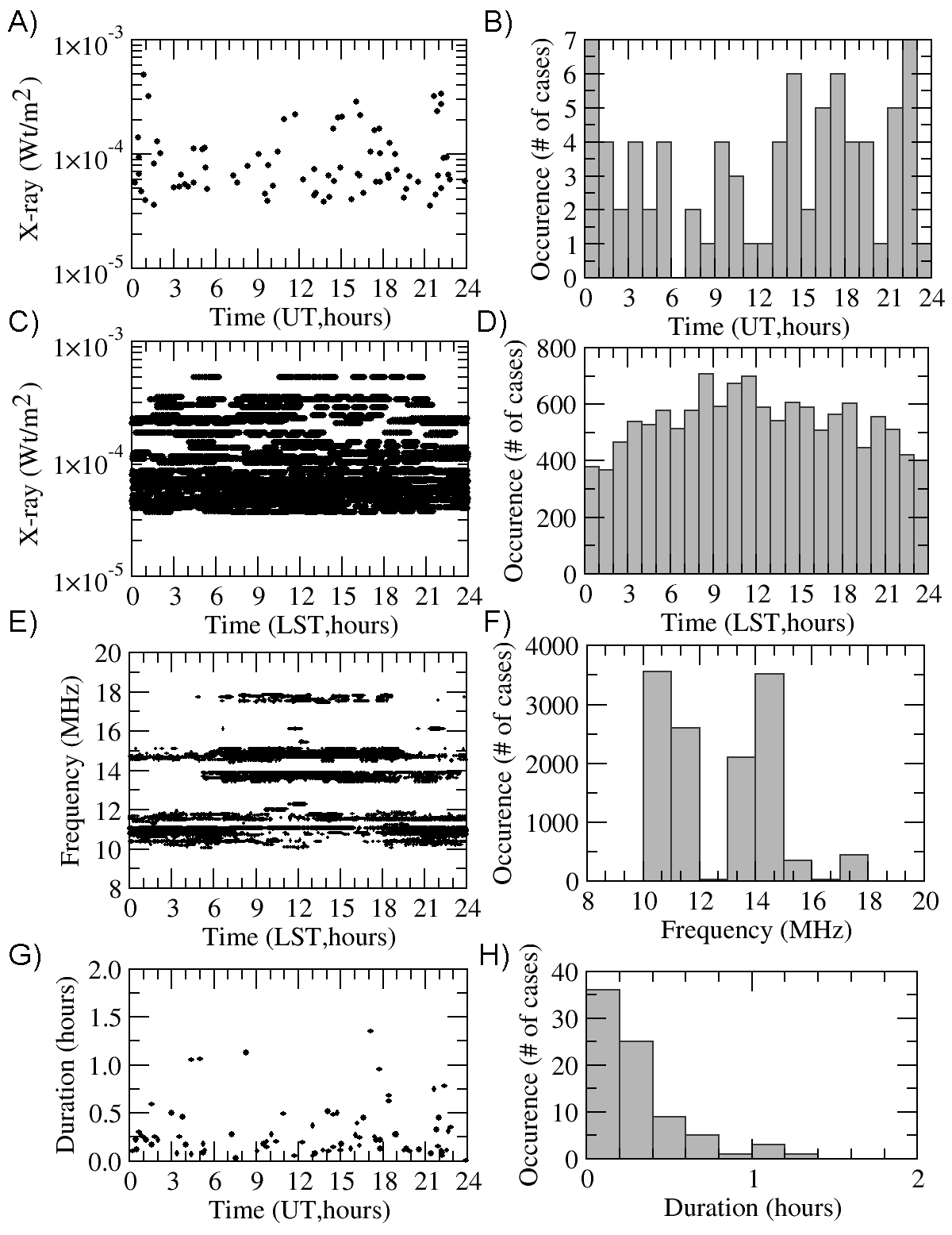}

\protect\caption{(A, C) - the distributions of x-ray solar flares analyzed in the paper
over maximal amplitude, as a function of the universal time (A) and
as a function of the local solar time (C), and (B, D) - solar flares
maximal amplitudes as functions of the universal time (B) and the
local solar time (D). E) - distribution of radar operating frequencies
during solar flares, as a function of the local solar time F) - statistical
distribution of radar operating frequencies during solar flares. E)-F):
distributions of flare duration (at $e^{-1}$ level): E) - as function
of UT, F) - statistical distribution.}
\label{fig:flare_distr}
\end{figure}

The distributions of flare maximal amplitudes and flare maximum moments are shown in
Fig.\ref{fig:flare_distr}. The distributions are made separately over
the universal time (Fig.\ref{fig:flare_distr}A) and over the local
time for each beam of each radar (Fig.\ref{fig:flare_distr}B ). The
local time in this case corresponds to a point determined from two factors - the
radar beam direction (most of the radars have 16 directions in their field-of-view),
and the distance from the radar to dead zone (or groundscatter distance)
\textasciitilde{} 1500 km. The dead zone distance is chosen due to
signal from a source can not approach a receiver by sky-wave mechanisms
at distances smaller than this. The approximation is very rough, since
the distance of the dead zone can vary from 500 to 3000 km, depending
on the background ionospheric conditions and the radar operating frequency.
However, as further analysis shows, our approximation allows us to obtain
sufficiently convincing statistical results. 

Therefore for each radar, each beam and each flare the daily dependence 
of the noise is collected.
As one can see in Fig.\ref{fig:flare_distr}C-D, the amount of the data collected by this way 
becomes quite large (the average number of experiments studied
for each hour is about 300) and pretty complete (the minimal number
of experiments for each hour is \textasciitilde{} 250).

The distribution of radar operating frequencies during the investigated
solar flares is shown in Fig.\ref{fig:flare_distr}E-F. From the Fig.\ref{fig:flare_distr}F
one can see that radars operated mostly in two bands: 10-12 MHz and
13-15 MHz, with nearly  absence of significant daily dependence 
of distributions (Fig.\ref{fig:flare_distr}E).
Thus, the data analyzed by us in this paper relate mostly to operating
frequencies 10-15 MHz.

Thus, we collected the statistics of observations of 80 solar flares
in more than 150 spatial points each. As a result we collected about
13000 daily measurements for statistical analysis. 
In Fig.\ref{fig:flare_distr}E-F is shown distribution of solar flare 
duration calculated at $e^{-1}$ level from maximal intensity.
As one can see from Fig.\ref{fig:flare_distr}E-F, the duration of
each of the flares does not exceeds 1.5~hour, with an average value about 20~min. 
This allows us to differ flare effects and daily noise dynamics by filtering.
Also it allows us to use standard radar temporal resolution mode 1-2min for 
investigating the x-ray flares effects.

From the pairs date-radar we excluded the following experiments, during which
the observed noise is too low or too high: 2015-03-09(WAL);
2015-03-09(BKS); 2015-03-10(WAL); 2015-03-10(BKS); 2015-03-11(BKS);
2015-03-11(WAL); 2014-04-18(FHE); 2014-04-18(FHW); 2015-08-22(CVW);
2015-06-22(FHE); 2015-08-24(CVW). Low and high noise can identificate 
that some not regular experiments of radar modes are carried out.

\begin{figure}
\includegraphics[scale=0.13]{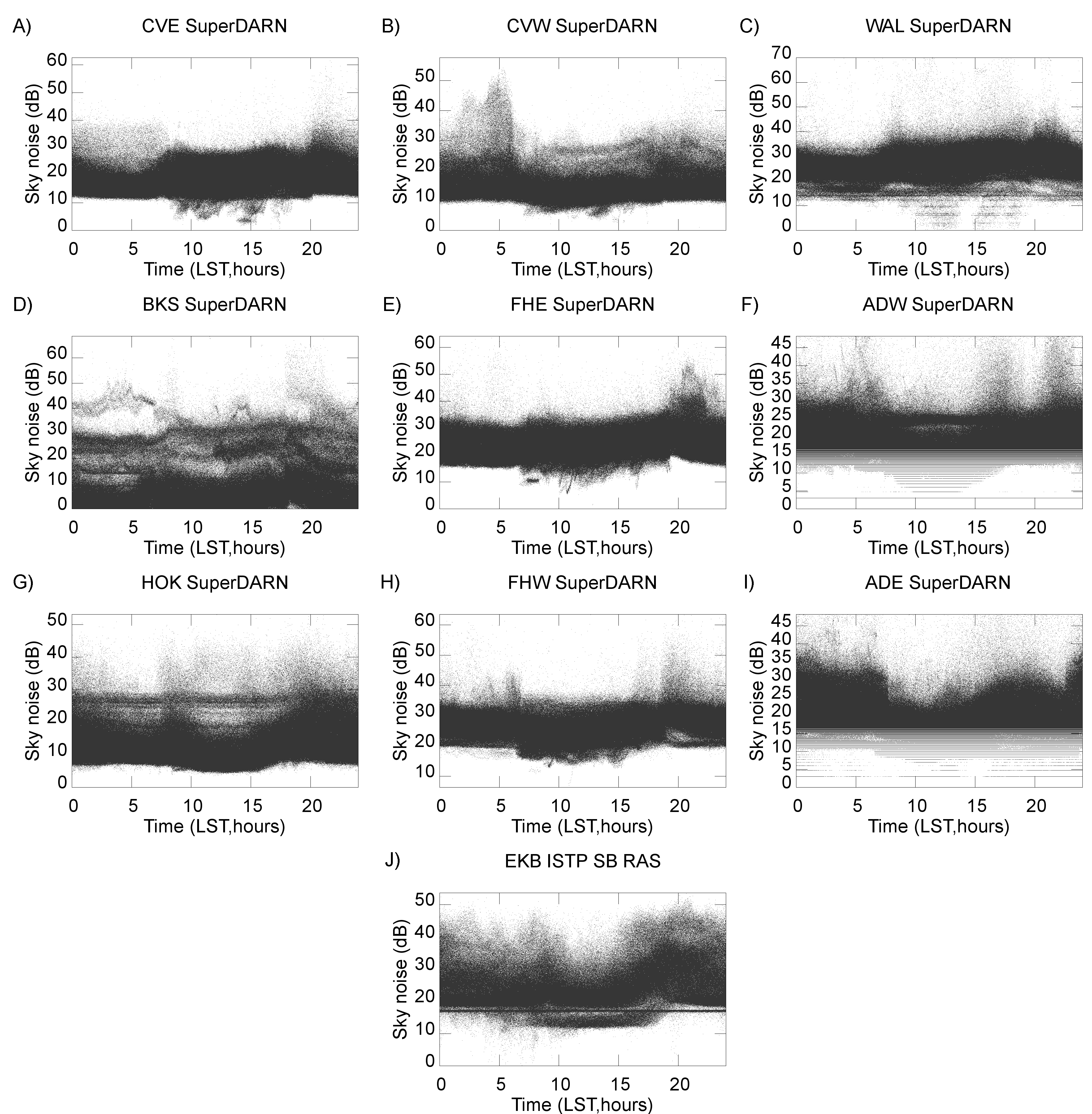}
\label{fig:noise_distrib}\protect\caption{Noise distribution for different radars averaged over all the beams
and over all the investigated days as a function of local solar time}
\end{figure}

In Fig.\ref{fig:noise_distrib} shown noise distribution
at each radars accumulated over all the beams and over all the investigated
days as a function of local solar time. As one can see, the noise
dynamics differs from one radar to another and has significant dynamics
and mean square variation (the variations of the noise intensity exceed 20-25dB). 
This can be caused by a set of factors: the changes of operating frequencies 
at each radar, different location and anthropogenic neighborhood,
different radar hardware.
High noise dynamics and variability 
do not allow us to use average or median values of the noise, calculated over all the experiments 
to accurately investigate solar flares effects. So to analyze the flare effect one should remove
background noise dynamics from the data for each day and each beam separately. To do this we
use the following algorithm.

To detect the effects of solar flares a correlation and regression
analysis was performed between the diurnal variation of the noise
and the flare intensity variation. The same was done for each day,
for each radar and for each radar beam. Since the intensity of the
received signal depends on the absorption exponentially \cite{Budden_1985},
we analyzed the relationship between the intensity of solar radiation
in $Wt/m^{2}$ and the noise intensity in dB. Changes of radar sounding
modes (in particular, the radar operating frequency or noise frequency)
for each radar was not taken into account to increase the time resolution
of the radar data.

To exclude the observed slow daily dynamics of sky noise, the slow varying component
of the noise level was removed from the noise intensity $N(t)$ using
an infinitive impulse response (IIR) filter with the impulse response
$e^{-t/\Delta T}$ (where $\Delta T$ is about 1 hour, exceeding everage flare duration). 
The resulting function $Y(t)$ is a relatively fast variation of the noise level
with periods of less than 1-2 hours:

\begin{equation}
Y(t)=10\left\{ lg(N(t))-\frac{1}{\Delta T}\int_{-\infty}^{t}lg(N(\tau))e^{-(t-\tau)/\Delta T}d\tau\right\} \label{eq:Y_form}
\end{equation}

To estimate the correlation between filtered noise ($Y(t)$) dynamics and x-ray intensity ($X(t)$) dynamics, the Pearson correlation
coefficient $R$ was used:

\begin{equation}
R=\frac{\int_{T_{0}-2\Delta T}^{T_{0}+2\Delta T}\left(X(t)-\left\langle X\right\rangle \right)\left(Y(t)-\left\langle Y\right\rangle \right)dt}{\sqrt{\int_{T_{0}-2\Delta T}^{T_{0}+2\Delta T}\left(X(t)-\left\langle X\right\rangle \right)^{2}dt\int_{T_{0}-2\Delta T}^{T_{0}+2\Delta T}\left(Y(t)-\left\langle Y\right\rangle \right)^{2}dt}}\label{eq:R_form}
\end{equation}

\begin{equation}
\begin{array}{c}
\left\langle X\right\rangle =\frac{1}{4\Delta T}\int_{T_{0}-2\Delta T}^{T_{0}+2\Delta T}X(t)dt\\
\left\langle Y\right\rangle =\frac{1}{4\Delta T}\int_{T_{0}-2\Delta T}^{T_{0}+2\Delta T}Y(t)dt
\end{array}\label{eq:XY_avrg}
\end{equation}

The Pearson correlation coefficient $R$ was calculated
over the period +/- 2h from the moment $T_{0}$ of the maximal flare
intensity. The correlation coefficient $R$ is within $[-1..1]$ and
defines the degree of correlation between the noise intensity and
the intensity of solar radiation during the flare. For high values $|R| > 0.5$, 
the correlation is significant and the two characteristics can
be considered as correlating well.

To study the proportionality of the logarithm of noise and the intensity
of the X-ray radiation during the flare, a regression analysis was
performed for each of the 13000 diurnal observations. According to
the method of least squares, the proportionality coefficient between
the variations of noise intensity and the variations of flare intensity
is defined as:

\begin{equation}
Y(t)\approx AX(t)+Y_{0}\label{eq:regression}
\end{equation}

where 

\begin{equation}
A=\frac{\int_{T_{0}-2\Delta T}^{T_{0}+2\Delta T}(X(t)-\left\langle X\right\rangle )(Y(t)-\left\langle Y\right\rangle )dt}{\int_{T_{0}-2\Delta T}^{T_{0}+2\Delta T}\left(X(t)-\left\langle X\right\rangle \right)^{2}dt}\label{eq:A}
\end{equation}

and $Y_{0}$ - some constant, unimportant for future analysis.

The resulting value of $A$ is regression coefficient between the variations 
of the noise intensity and the intensity of the x-ray flare.

\section{Results}

\begin{figure}
\includegraphics[scale=0.5]{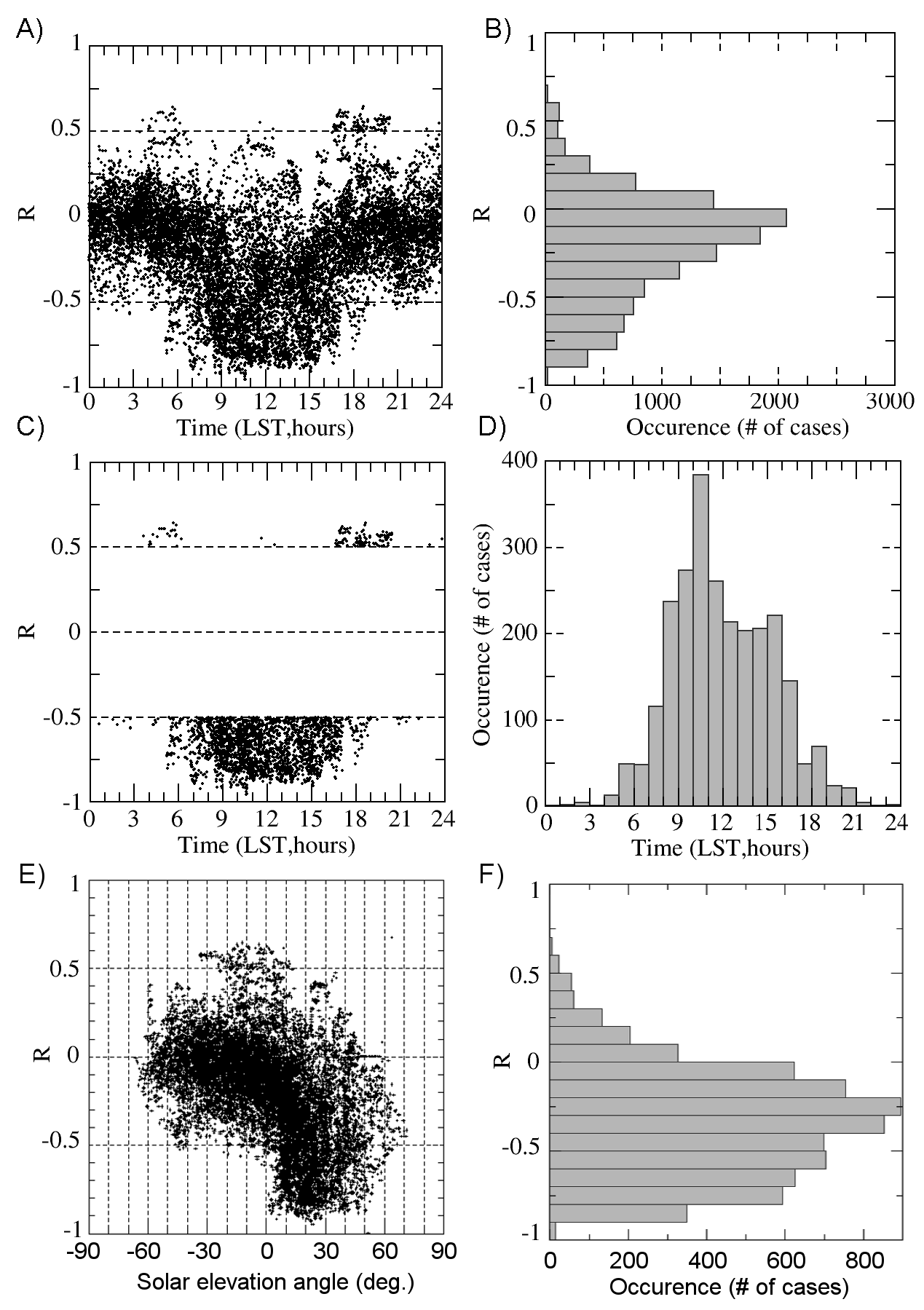}

\protect\caption{A) - Pearson correlation coefficients between the measured sky noise
and x-ray flare intensity for different experiments as a function
of local solar time B) - Statistical distribution of Pearson correlation
coefficients. C) - Experiments with significant Pearson correlation
($|R| > 0.5$) between measured sky noise and x-ray flare intensity
as a function of local solar time. D) Local solar time distribution
of cases with significant Pearson correlation ($|R|>0.5$)
E) Dependence of Pearson correlation coefficient on solar elevation angle
F) Distribution of Pearson correlation coefficient for positive elevation angles
}
\label{fig:corr_results}
\end{figure}

Based on the selected set of \textasciitilde{}13000 daily
measurements by 10 radars, as well as based on GOES X-ray measurements
in the range 0.1-0.8 nm during 80 flares, we performed correlation
and regression analysis according to approach (\ref{eq:Y_form}-\ref{eq:A}).
The resulting values of the
correlation coefficient $R$ are analyzed as a function of local solar
time LST (LST was determined from the position of the radar, the
direction of the beam of the radar antenna pattern and for effective attenuation range 1500km). 
The results of the correlation analysis are shown in Fig.\ref{fig:corr_results}.
From Fig.\ref{fig:corr_results}A-B one can see that the most part
of experiments (\textasciitilde{}72\%) produces the correlation coefficient $-0.5 \le R \le 0.5$. 
Usually this can be interpreted as statistically
not correlated variations of the sky noise and intensity of solar
x-ray flares. Nevertheless, from the statistical distribution of the correlation
coefficients as a function of local time (shown in Fig.\ref{fig:corr_results}B)
one can see significant asymmetry and negative offset of the correlation
coefficient ($R<0$ in 68\% cases) and the average value of correlation coefficient $R$ is about
$-0.21\pm0.29$. This can be interpreted as a sign of predominant attenuation
of HF sky-noise during x-ray solar flares.

The observed statistical effect confirms our preliminary results
\cite{Berngardt_2017}, used for the research motivation.

In Fig.\ref{fig:corr_results}D one can see that the correlation coefficient
exceeds the limits {[}-0.5..0.5{]} mostly during daytime 07-19LST.
Daytime observations (07-19LST) correspond to about 93\% of all the observations
with $|R|>0.5$.
The median moment of observations with $|R|>0.5$ corresponds
to \textasciitilde{}11:30LST with quartiles at \textasciitilde{}09:30LST and \textasciitilde{}14:30LST. 
This means that most of the cases with strong correlaion 
between sky-noise and x-ray intensity corresponds to daytime, and the effect 
is centered near the solar noon. This allows us to suggest that the effect is really 
connected with the Sun activity and arises in the radar vicinity.

In Fig.\ref{fig:corr_results}E shown dependence of Pearson correlation coefficient on solar elevation angle. As one can see,
during daytime (54\% of cases) the correlation significantly intensify. In Fig.\ref{fig:corr_results}E shown the distribution 
of Pearson correlation coefficient for positive elevation angles (i.e. during daytime). As one can see during daytime the strong 
negative correlation of the effect is more clear that at whole statistics (Fig.\ref{fig:corr_results}B).

It should be noted that most part of the daytime data (67\%) is poorly 
correlated with the solar flare
($|R| <0.5$). Nevertheless, the negative correlation ($R<0$) is observed
in 88.3\% of all the daytime cases, and $R <-0.5$ is observed
in 33\% of all the daytime cases. 
This indicates that daytime noise is attenuated by x-ray solar flare, but 
there can be a significant dependence of the magnitude of the effect on 
other parameters, for example, from the initial level of background noise, 
from the geometry of terrestrial noise sources, as well as from the local 
atmospheric and ionospheric conditions near the radar.

\begin{figure}
\includegraphics[scale=0.5]{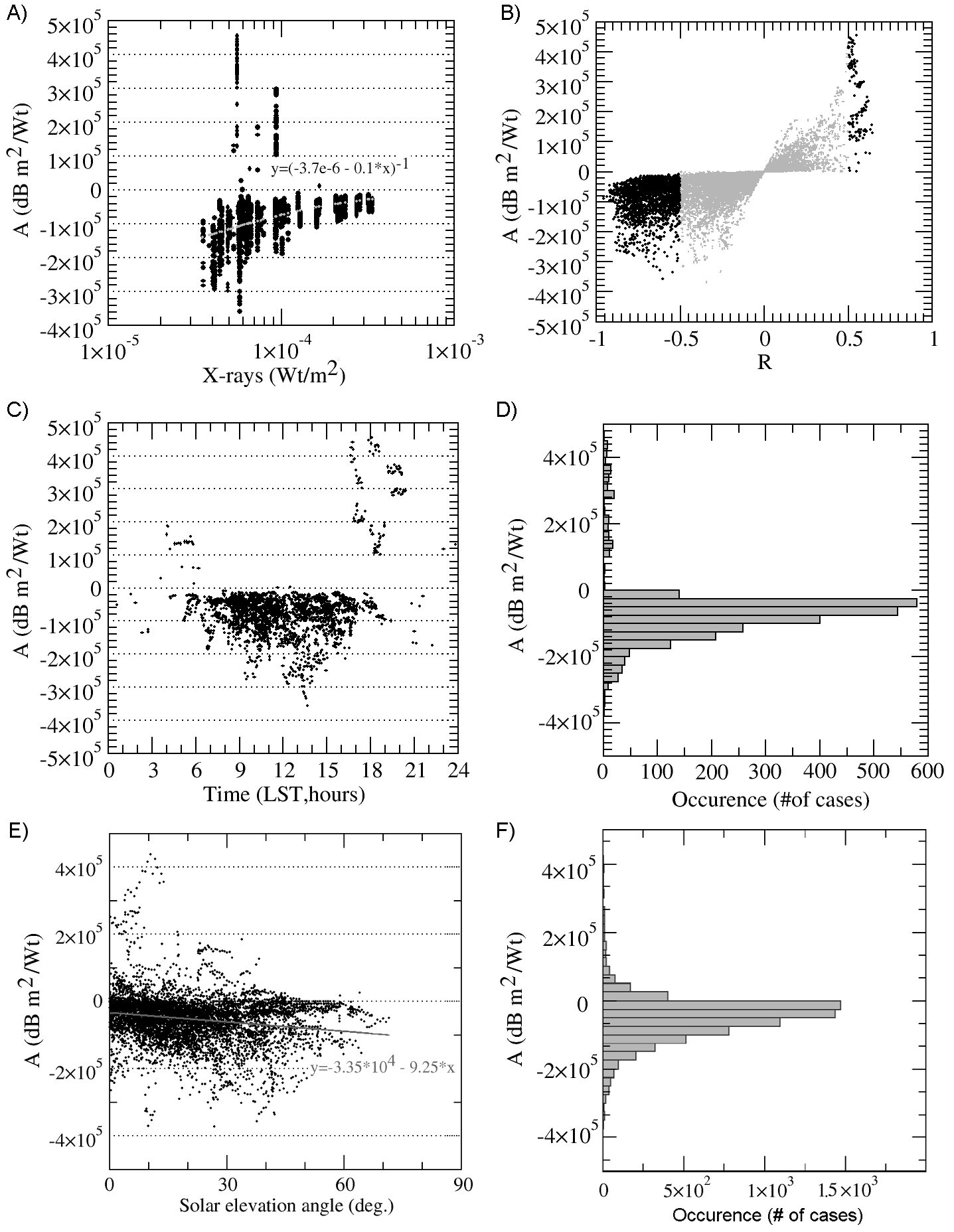}
\protect\caption{(A) is the regression dependence between the intensity of the effect
in the HF sky noise and the x-ray intensity of the flare for the
good correlating experiments ($|R|>0.5$); (B) - dependence of regression coefficient $A$ on
correlation coefficient $R$;  (C-D) is the statistics of regression
coefficient A (\ref{eq:A}): its dependence on the local time (C)
and its statistical distribution (D);
(E)  - dependence of regression coefficient $A$ on solar elevation angle;
(F) - statistical distribution of regression coefficient $A$ for positive solar elevation angles.
}

\label{fig:regression_goes_noise}
\end{figure}

In Fig.\ref{fig:regression_goes_noise}A shown the dependence of attenuation of sky noise 
on the x-ray intensity for good correlation ($|R|>0.5$). One can see that there is apparently no linear
regression between noise level attenuation and x-ray flare intensity.
But there is an obvious tendency to increase noise attenuation during
the daytime flares (Fig.\ref{fig:regression_goes_noise}B). The average
regression coefficient for $|R|>0.5$ is about $(-6.9\pm10.9) \cdot10^{4}[dB\cdot m^{2}/Wt]$.
The average
regression coefficient for $|R|>0.7$ is about $(-9.9\pm9.5) \cdot10^{4}[dB\cdot m^{2}/Wt]$.

The big mean square error shows that distribution is too wide to be estimated it by mean value.
The distribution of the regression coefficients A and their daily
dependence are shown in Fig.\ref{fig:regression_goes_noise}B-C. As
one can see, the distribution is wider than average value, but most
part of it is negative. This allows us to conclude that during well correalting ($|R|>0.5$)
experiments the noise attenuates during x-ray flares. The distribution of the
regression coefficient $A$ also shows the negative median regression coefficient for ($|R|>0.5$):
$A \approx -6.9\cdot10^{4}\ [dB\cdot m^{2}/Wt]$ with quartiles $-4.2\cdot10^{4}, -11.2\cdot10^{4} [dB\cdot m^{2}/Wt]$. 
For ($|R|>0.7$) median value reaches
$A \approx -8.0\cdot10^{4}\ [dB\cdot m^{2}/Wt]$ with quartiles $-5.8\cdot10^{4}, -12.4\cdot10^{4} [dB\cdot m^{2}/Wt]$. 

In Fig.\ref{fig:regression_goes_noise}E shown dependence of regression coefficient $A$ on solar elevation angle during daytime.
As one can see there is a slight increase of regression coefficient with increase of the solar elevation angle.
This can be explained by an increase in the ionization level with an increase in the solar elevation angle 
and explains the increase of regression coefficient at noon (Fig.\ref{fig:regression_goes_noise}B).

In Fig.\ref{fig:regression_goes_noise}F shown the statistical distribution of regression coefficient $A$ 
during daytime - for positive solar elevation angles. Average regression coefficient $A$ during daytime 
is $-5.3\cdot10^{4}[dB\cdot m^{2}/Wt]$ with mean square error $7.5\cdot10^{4}[dB\cdot m^{2}/Wt]$. Median value 
of $A$ is $-4.4\cdot10^{4}[dB\cdot m^{2}/Wt]$ with quartiles $-1.5\cdot10^{4}[dB\cdot m^{2}/Wt]$, $-8.6\cdot10^{4}[dB\cdot m^{2}/Wt]$.

At nighttime the average $R$ is about -0.04, and median $A$ is $-6.3\cdot10^{3}[dB\cdot m^{2}/Wt]$ i.e. there is almost no 
signifficant correlation between solar xray flare and noise attenuation at night. The nighttime regression coefficient $A$ 
is about ten times smaller than daytime $A$, that also shows the greater importance of the daytime effect than the nighttime effect.

Let us interpret this. Solar xray radiation mostly effects at illuminated part of the Earth's ionosphere. 
If the far-range noise sources makes the main contribution 
into the sky-noise, measured by the radar, the effect should not significantly depend on radar local solar time.
But according to experimental data (Fig.\ref{fig:regression_goes_noise}E) the strong effect exists only when the radar 
is at dayside. 
This allows us to exclude from consideration the sources of the noise, located too far from the radar, and suppose 
that the main contribution to the radar noise is made by the sources located near the radar 
(we expect about 1500km).

\section{Discussion and conclusion}

Our statistical analysis of the Pearson correlation coefficient
($R$) and the regression coefficient ($A$) of 80 x-ray flares at 10 mid-latitude HF radars, about 16 directions for each 
(which is equivalent to about 13000 individual experiments)
shows that there is a certain tendency (88.3\% of daytime experiments) to attenuate daytime HF noise
level during x-ray solar flares, about 34\% of the attenuation profile correlates well ($|R|>0.5$) with x-ray intensity profile. 
So the attenutation of the daytime HF noise during solar xray flates can be interpreted as a regular effect.
During nightime the sky-noise attenutation is observed in 62\% of cases, and in the most of the cases (97.6\%) the 
attenuation profile is poorly correlating with x-ray intensity profile ($|R|<0.5$).
So the nighttme effects in comparison with daytime effect can be neglectable.

\begin{figure}
\includegraphics[scale=0.12]{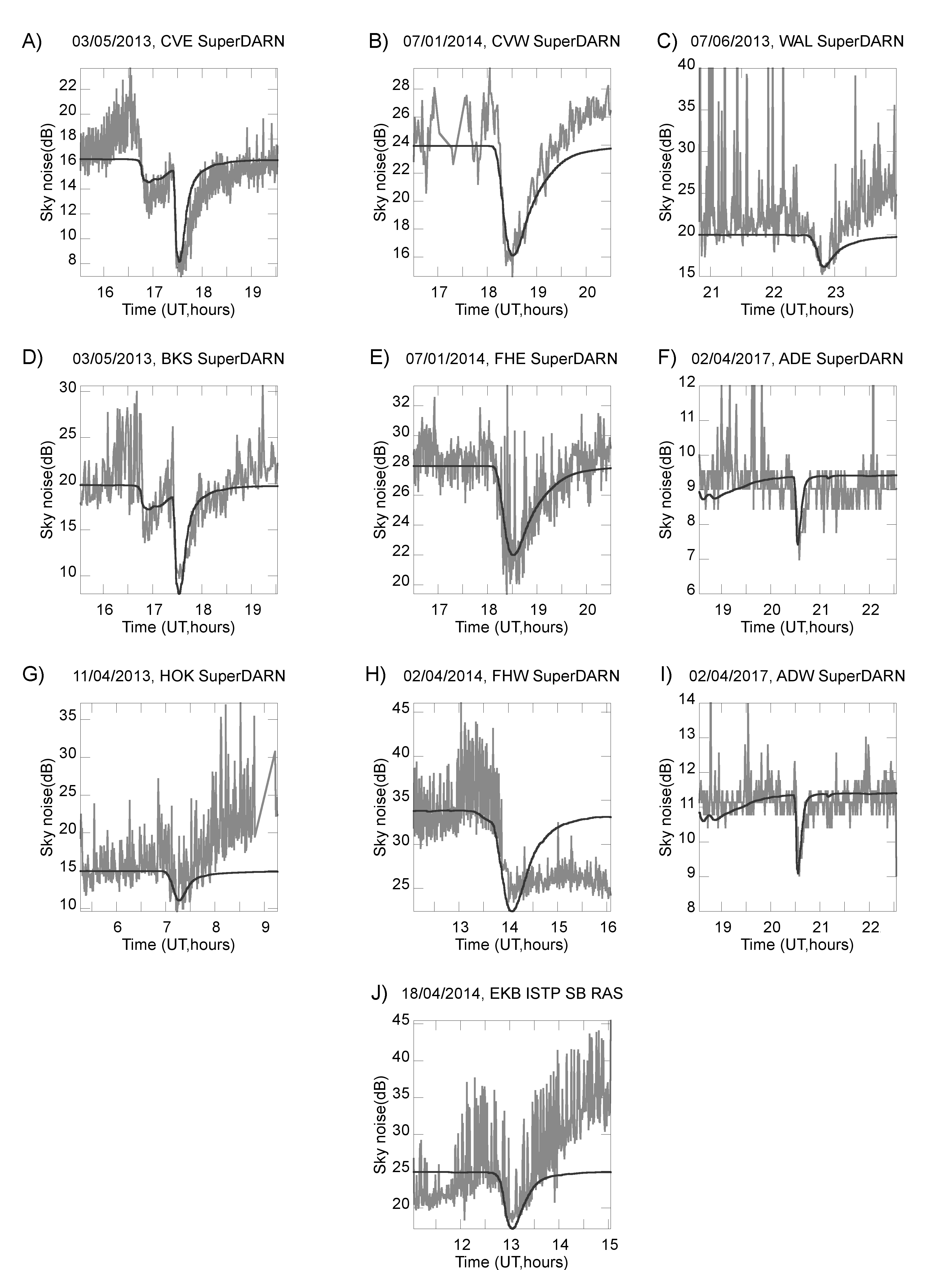}
\protect\caption{Examples of daytime sky noise dynamics at different radars during
different x-ray solar flares (for cases with good Pearson correlation
coefficient $R<-0.5$). Red line - GOES x-ray intensity (normalized),
green line - sky noise dynamics.}

\label{fig:noise_examples_SD}
\end{figure}

Fig.\ref{fig:noise_examples_SD}
shows illustrative examples of noise attenuation during solar flares
at various radars for good Pearson coefficient ($R<-0.5$). The figure 
illustrates a good correlation of the temporal dynamics of X-ray 
radiation and the dynamics of sky noise level, which confirms with 
the statistically observed tendency to high anticorrelation between them
(shown in Fig.\ref{fig:corr_results}-\ref{fig:regression_goes_noise}).

\begin{figure}
\includegraphics[scale=0.22]{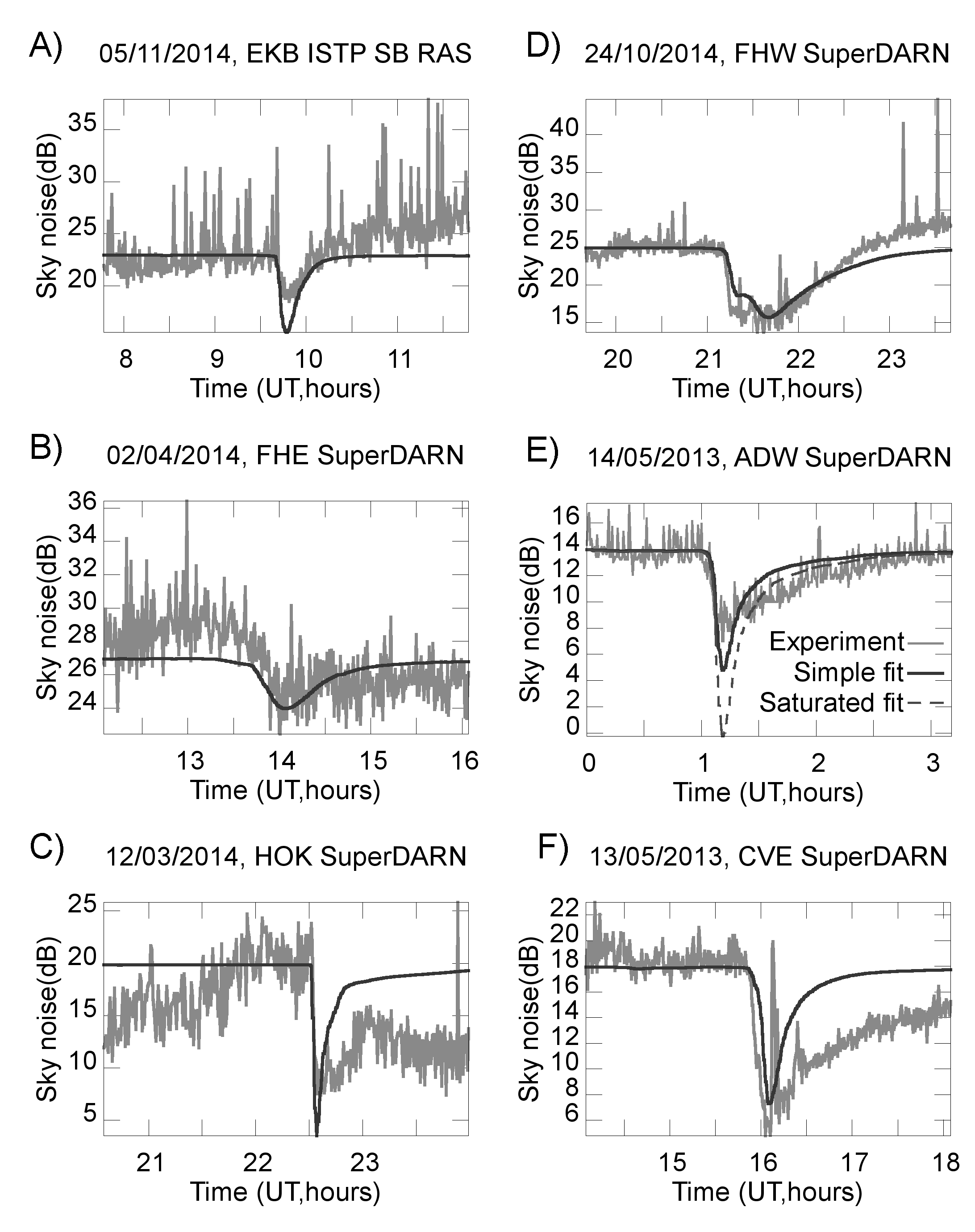}
\protect\caption{Sources of errors in calculation of A. A-C) - Sky noise dynamics effect;
D-F) - saturation effect; E) - illustration of saturation effect at
ADW 14/05/2013 data example: gray line - experimental data; black
line - fitted x-ray intensity without taking into account presence
of saturation effect; dashed line - fitted x-ray intensity with
taking into account presence of saturation effect.}
\label{fig:err_sources}

\end{figure}

As it can be seen from Fig.\ref{fig:noise_examples_SD}F,I, the minimal
self-noise of the receivers usually is about 1-10dB (this corresponds
to the lowest 3 bits of digital signal). When sky noise intensity
becomes lower than 10dB level it becomes hard to investigate the noise
dynamics. In Fig.\ref{fig:err_sources} the possible sources
of errors in calculation of regression coefficient $A$ are illustrated.
They are: the own dynamics of sky noise intensity is not taken into account correctly
(Fig.\ref{fig:err_sources}A-C); the presence of additional noises, not
affected by flare and too powerful effect causing saturation effect
during flare maximum (Fig.\ref{fig:err_sources}C,D-F). In Fig.\ref{fig:err_sources}E
the effect of saturation is shown more clear: the x-ray intensity
is fitted without taking into account saturation effect (black dashed
line) and with taking into account the possibility of saturation effect (black solid
line). As one can see, the results are different, so taking into account saturation 
effect increase regression coefficient value $A$.

In Fig.\ref{fig:ill_geometry}I shown the proposed interpretation of
the observed effect, similar to well known explanation of HF absorbtion 
of radiowave signals \cite{Hunsucker_2002}. Decameter 
radiowaves from closely located sources
does not reflect from the ionosphere and does not attenuate in its
lower part (trajectory 1 in Fig.\ref{fig:ill_geometry}I). Decameter
radiowaves outside the dead zone reflects from the ionosphere and
passes through the ionospheric D-layer twice, having substantial absorption
during the x-ray solar flares due to additional D-layer ionization
(trajectory 2 in Fig.\ref{fig:ill_geometry}I). The direct solar radiation
in the HF range to the receiver falls or passes depending on the ionospheric
conditions - when the operating frequency is below the plasma frequency
f0F2 - it is reflected back, and when the operating frequency is higher
than f0F2 - it passes (trajectory 3 in Fig.\ref{fig:ill_geometry}I).
Most often (Fig.\ref{fig:corr_results}B,\ref{fig:regression_goes_noise}C)
the combination of all these mechanisms is observed as attenuation
of the HF sky noise, and this attenuation in the logarithmic scale
nearly repeats the shape of the x-ray intensity during the flare (Fig.\ref{fig:noise_examples_SD}).
Dependence of noise effects on the radar beam (azimuth,Fig.\ref{fig:ill_geometry}A-C,E-G) allows 
us to neglect the cosmic noise role in the effect.
This indicates the predominance of the mechanism of over-the-horizon noise
propagation from terrestrial (anthropogenic) sources over the effects
of cosmic noise and ground-wave noise in the investigated frequency
range of 10-15 MHz during solar x-ray flares. Thus, it can be assumed
that in this case the influence of anthropogenic noise sources
is significant and should be taken into account when interpreting
the data.

The median regression coefficient $A$ between HF sky noise and 0.1-0.8nm
GOES x-ray intensity during daytime is about $-4.4\cdot10^{4}[dB\cdot m^{2}/Wt]$.
The significant contrast between regression coefficient $A$ and constant value (Fig.\ref{fig:regression_goes_noise}A)
can be associated with many effects, that requires future investigations
and corrections - mostly these are: HF sky noise dynamics;  non-linear regression dependence including saturation
effect due to powerful flares; additional noise, not affected by
ionosphere. Another reason for possible errors in calculations of $A$ is the irregular
changes of the radar operational frequency in the standard scanning
mode. This can also lead to additional variations in the amplitude
of the noise and to increase of errors.
Also there is shown a slight increase of regression coefficient $A$ with solar elevation angle. This 
can be explained by increase of ionization with increase of solar elevation angle.

Thus, in the paper shown that daytime short-wave noise during solar flares
in 88.3\% of cases attenuates, which can be explained by terrestrial
(anthropogenic) source of HF sky noise. As one can see the regression
coefficient $A$ between x-ray intensity and noise intensity in logarithm
scale is useful parameter for describing the ionospheric absorption
effects during x-ray solar flares. Therefore, HF sky noise level dynamics
can be used to monitor ionospheric effects during x-ray solar flares
by mid-latitude decameter radars, and these radars can be used as
a kind of mid-latitude riometers to investigate reaction of D-layer
to x-ray solar flares.

\section{Acknowledgments}

In the paper we used the data of EKB ISTP SB RAS, operating under financial
support of FSR Program II.12.2.3. 
The SuperDARN Hokkaido East is the property of Nagoya University and
its construction/operation has been funded by the Ministry of
Education, Culture, Sports, Science and Technology of Japan.
O.B. is supported by Presidium RAS Program \#31 (Project Number: \#0344-
2014-0022), N.N. is supported by Japan Society for the Promotion
of Science (JSPS), "Grant-in-Aid for Specially Promoted Research"
(Project Number: 16H06286).

The data of the SuperDARN radars
were obtained using the DaViT (https:// github.com/ vtsuperdarn/ davitpy),
the EKB ISTP SB RAS radar data are the property of the ISTP SB RAS,
contact Oleg Berngardt (berng@iszf.irk.ru). The authors are grateful
to NOAA for providing the GOES data. The authors are grateful to all
the developers of the DaViT system, in particular K.Sterne, N.Frissel,
S. de Larquier and A.J.Riberio, as well as to all the organizations
supporting the radars operation.


\end{document}